\documentstyle[12pt]{article}
\newcommand{\bm}[1]{\mbox{\boldmath $#1$}}

\newcommand{\ol}{\overline}
\newcommand{\be}{\begin{equation}}
\newcommand{\ee}{\end{equation}}
\newcommand{\ba}{\begin{eqnarray}}
\newcommand{\ea}{\end{eqnarray}} 
\newcommand{\lb}[1]{\label{#1}}  
\newcommand{\bb}[1]{\bibitem{#1}}
\newcommand{\E}{{\cal E}}
\newcommand{\e}{{\rm e}}

\begin{document}
\begin{titlepage}
\setcounter{page}{1}
\title{Self--gravitating cosmic rings}
\author{
\bigskip
G\'erard Cl\'ement\thanks{E-mail: gecl@ccr.jussieu.fr.}\\ 
\small Laboratoire de Gravitation et Cosmologie Relativistes,\\
\small Universit\'e Pierre et Marie Curie, CNRS/UPRESA 7065,\\
\small Tour 22-12, Bo\^{\i}te 142, 4 place Jussieu,
75252 Paris cedex 05, France 
}
\bigskip
\date{September 1, 1998}
\maketitle

\begin{abstract}
The classical Einstein--Maxwell field equations admit static horizonless
wormhole solutions with only a circular cosmic string singularity. We show
how to extend these static solutions to exact rotating asymptotically flat
solutions. For a suitable range of parameter values, these solutions
describe charged or neutral rotating closed cosmic strings, with a
perimeter of the order of their Schwarzschild radius.
\\ [15mm]
{\em PACS}: \hspace{5mm} 04.20 Jb \hspace{3mm}  04.40 Nr \hspace{3mm}
98.80 Cq
\end{abstract}

\bigskip
\bigskip

\end{titlepage}

The Einstein--Maxwell field equations coupling gravity to electromagnetism
admit outside sources a variety of stationary axially symmetric solutions,
among which the Kerr--Newman family of solutions \cite{newman} depending
on three parameters $M$, $Q$ and $J$ which, from the consideration of the
asymptotic behaviour of these field configurations, may be identified as
their total mass, charge, and angular momentum. A fourth physical
characteristic, the total magnetic moment $\mu$, is related to these three
parameters by the same ``anomalous'' relation
\be \lb{gyro}
g \equiv 2\frac{M\mu}{JQ} = 2
\ee 
as in the case of an elementary particle such as the electron. It would be
tempting to interpret these classical configurations as the fields
generated by an isolated elementary particle, were it not for the
numerical values of these parameters. The Kerr--Newman metrics correspond
to regular black--hole spacetimes if
\be \lb{bh}
M^2 \ge Q^2 + a^2,
\ee
where\footnote{We use gravitational units $G = 1$, $c = 1$.} $a \equiv
J/M$. However in the case of elementary particles $|J| \sim m_P^2$ (where
$m_P = (\hbar c/G)^{1/2}$ is the Planck mass) and $|Q| \sim m_P$, so that
$$
|a|/|Q| \sim |Q|/M \sim m_P/m_e \sim 10^{22}
$$
(where $m_e$ is the electron mass), and therefore
\be \lb{part}
M^2 < Q^2 + a^2\,.
\ee
So the field configurations
generated by elementary particles are not of the black hole type, but
exhibit naked singularities, violating the cosmic censorship hypothesis
\cite{censor}. For this reason, it is generally believed that there is no
viable classical model for elementary particles (except possibly for
neutral spinless particles) in the framework of Einstein's general relativity.

In the charged spinless case $J = 0$, the relation $|Q|/M \sim m_P/m_e$
tells us that electromagnetism is preponderant and that the point
singularity in the spherically symmetric metric originates from that of
the Coulomb central field. In classical models of (spinless) charged point
particles, this singularity leads to the divergence of the particle's
self--energy, and the situation becomes much worse in quantum field
theory. These divergences may be tamed in quantum string theories where
the elementary objects are no longer zero--dimensional (point particles)
but one--dimensional (strings). Similarly, one expects that the
divergences in a classical field theory with line sources will be less
severe than with point sources. String--like objects occur as classical
solutions to field theories with spontaneously broken global or local
symmetries. Such symmetry--breaking transitions are believed to have
occurred during the expansion of the universe, leading to the formation of
large, approximately straight cosmic strings \cite{cs1}. The long--range
behavior of the metric generated by a straight cosmic string is given by
an exact stationary solution of the vacuum Einstein equations with a line
source carrying equal mass per unit length and tension \cite{cs2}. In the
case of closed cosmic strings, or rings, this tension will cause the
string to contract, precluding the existence of stationary solutions,
unless the tension is balanced by other forces (superconducting cosmic
strings
\cite{scs}, vortons \cite{vort}). Another possibility has been advocated
by Bronnikov and co--workers \cite{kb}, that of ring wormhole solutions to
multi--dimensional field models. In the case of these static solutions, the
gauge field energy--momentum curves space negatively to produce a
wormhole, at the neck of which sits a closed cosmic string, which cannot
contract because its circumference is already minimized.

In this Letter, we first rederive such static ring wormhole solutions to
the Einstein--Maxwell field equations. Then, using a recently proposed
spin--generating method \cite{kerr}, we construct from these static
solutions new rotating Einstein--Maxwell ring solutions with only a cosmic
ring singularity.  These solutions depend on four parameters, the values
of which can be chosen such that the elementary particle constraints
(\ref{gyro}) (slightly generalized to $g \approx 2$) and (\ref{part})  are
satisfied. However, it then turns out that for the ``elementary'' orders
of magnitude $|J| \sim Q^2 \sim m_P^2$, the mass of these objects cannot
be small, but is also of the order of the Planck mass. We show that, in
the case of large quantum numbers ($|J| \gg m_P^2$), a subclass of these
solutions describes macroscopic charged or neutral rotating cosmic rings,
also satisfying the elementary particle constraint (\ref{part}), but with 
$g \neq 2$.

Under the assumption of a timelike Killing vector field $\partial_t$, the
spacetime metric and electromagnetic field may be parametrized by 
\ba\lb{stat1}
ds^2 & = & f\,(dt - \omega_i dx^i)^2 - f^{-1}\,h_{ij}\, dx^i dx^j\,,
\nonumber\\
F_{i0} & = & \partial_i v\,, \qquad F^{ij} = f\,h^{-1/2}\epsilon^{ijk}
\partial_k u\,,
\ea
where the various fields depend only on the three
space coordinates $x^i$. The stationary Einstein--Maxwell equations are
equivalent to the three--dimensional Ernst equations \cite{er}
\ba
f\nabla^2\E & = & \nabla\E \cdot (\nabla\E + 
2\ol{\psi}\nabla\psi)\,, \nonumber \\
f\nabla^2\psi & = & \nabla\psi \cdot (\nabla\E + 
2\ol{\psi}\nabla\psi)\,, \lb{ernst} \\
f^2R_{ij}(h) & = & {\em Re} 
\left[ \frac{1}{2}\E,_{(i}\ol{\E},_{j)} 
+ 2\psi\E,_{(i}\ol{\psi},_{j)}
-2\E\psi,_{(i}\ol{\psi},_{j)} \right]\,, \nonumber
\ea
where the scalar products and Laplacian are computed with the metric
$h_{ij}$, the complex Ernst potentials $\E$ and $\psi$ are defined by
\be
\E = f - \ol{\psi}\psi + i\chi\,, \qquad \psi = v + iu\,,
\ee
and $\chi$ is the twist potential 
\be\lb{twist}
\partial_i\chi = -f^2\,h^{-1/2}h_{ij}\,\epsilon^{jkl}\partial_k\omega_l 
+ 2(u\partial_i v - v\partial_i u)\,.
\ee
These equations are invariant under an SU(2,1) group of transformations
\cite{nk}. The class of electrostatic solutions ($\E$ and $\psi$ real)
depending on a single real potential can be reduced, by a group
transformation, to $\E = \E_0$ constant, which solves the first equation 
(\ref{ernst}). The form of the solution of the second equation (\ref{ernst})
depends on the sign of $\E_0$. Representative solutions are
\ba\lb{stat2}
\E_0 = -1\,, \qquad & \psi_0 = \coth(\sigma)\,, \qquad 
& f_0 = 1/\sinh^2\sigma \nonumber\\
\E_0 = 0\,, \qquad & \psi_0 = 1/\sigma\,, \qquad 
& f_0 = 1/\sigma^2 \\
\E_0 = +1\,, \qquad & \psi_0 = \cot(\sigma)\,, \qquad 
& f_0 = 1/\sin^2\sigma \nonumber 
\ea
where the potential $\sigma(\bm{x})$ is harmonic ($\nabla^2\sigma = 0$); other 
electrostatic solutions depending on a
single potential may be obtained from these by SU(2,1) transformations.
We note that the electric and gravitational potentials (\ref{stat2}) are 
singular for $\sigma = 0$ if $\E_0$ = --1 or 0, and for $\sigma = n\pi$ ($n$
integer) if $\E_0$ = +1. 

As we wish to obtain axisymmetric ring--like solutions, we
choose oblate spheroidal coordinates $(x,y)$, related to the familiar Weyl
coordinates $(\rho,z)$ by
\ba \lb{oblate1}
& \rho & = \nu \,(1+x^2)^{1/2}(1-y^2)^{1/2}\,, \nonumber \\
& z & = \nu \,xy\,.
\ea
In these coordinates, the three--dimensional metric $d\sigma^2 \equiv
h_{ij}dx^idx^j$
\be \lb{oblate2}
d\sigma^2 = \nu^2\,[\e^{2k}(x^2+y^2)(\frac{dx^2}{1+x^2} +
\frac{dy^2}{1-y^2}) + (1+x^2)(1-y^2)\,d\varphi^2\,]
\ee
depends on the single function $k(x,y)$. Now, following Bronnikov et al.\
\cite{kb}, we assume the harmonic potential $\sigma$ to depend only on the
variable $x$, which yields
\be
\sigma = \sigma_0 + \alpha \arctan x\,, \qquad \e^{2k} =
\left(\frac{1+x^2}{x^2+y^2}\right)^{\E_0\alpha^2}\,,
\ee
where $\sigma_0$ and $\alpha$ are integration constants. We note that the
reflexion $x \leftrightarrow -x$ is an isometry  for the
three--dimensional metric (\ref{oblate2}), which has two points at
infinity $x = \pm \infty$. The full four--dimensional metric (\ref{stat1})
is quasi--regular (i.e.\ regular except on the ring $x=y=0$, see below)
for $x \in R$ if
\ba\lb{reg1}
& |\sigma_0| > |\alpha|\pi/2 \quad & {\rm for}\;\;\E_0 = -1\;{\rm or}\;0 
\nonumber \\
& (n+|\alpha|/2)\pi < \sigma_0 < (n+1-|\alpha|/2)\pi \qquad  (|\alpha| < 1)
\quad & {\rm for}\;\;\E_0 = +1
\ea
for some integer $n$. If
these conditions are fulfilled, this metric describes a wormhole spacetime
with two asymptotically flat regions connected through the disk $x=0$
($z=0, \rho < \nu$). There is no horizon. The point singularity of the 
spherically symmetric (Reissner--Nordstr\"{o}m) solution is here
spread over the ring $x=y=0$ ($z=0, \rho=\nu$), near which the behavior of
the spatial metric 
\be
d\sigma^2 \simeq \nu^2\,[(x^2+y^2)^{1-\E_0\alpha^2}(dx^2+dy^2) + d\varphi^2]
\ee
is that of a cosmic string with deficit angle $\pi(\E_0\alpha^2-1)$, which
is negative in all cases of interest (it can be positive only for $\E_0 =
+1$, $|\alpha| > 1$, corresponding to a singular solution). This ring
singularity disappears in the limit of a vanishing deficit angle ($\E_0 =
+1, |\alpha| \to 1$), where the solution reduces to a
Reissner--Nordstr\"{o}m solution with naked point singularity.
The asymptotic behaviours of the gravitational and
electric potentials at the two points at infinity are those of particles
with masses and charges 
\be
M_{\pm} =
\mp\alpha\nu\frac{\psi_0(\pm\infty)}{\sqrt{f_0(\pm\infty)}}\,, \qquad 
Q_{\pm} = \pm\alpha\nu\,;
\ee
the three cases (\ref{stat2}) lead respectively to
$Q_{\pm}^2 < M_{\pm}^2$ for $\E_0 = -1$, $Q_{\pm}^2 = M_{\pm}^2$ for $\E_0
= 0$, and $Q_{\pm}^2 > M_{\pm}^2$ for $\E_0 = +1$.  The vanishing
of the sum of the outgoing electric fluxes at $x = \pm\infty$ shows that
the ring $x = y = 0$ is uncharged.

A case of special interest is $\E_0 = +1$, $\sigma_0 = \pi/2$,
corresponding to a symmetrical wormhole metric \cite{ch}.  The mass of
this particle $M_{\pm} = \alpha\nu\sin(\alpha\pi/2)$ does not depend on
the point at infinity considered, and is positive, even though the deficit
angle is negative.  For the physical characteristics of this particle to
be those of a spinless electron, we should take $|\alpha|\sim m_e/m_P$,
and $\nu \sim m_P^2/m_e$, of the order of the classical electron radius.

Now we proceed to generate asymptotically flat rotating solutions from
these static axisymmetric solutions, using the simple procedure recently 
proposed in \cite{kerr}. This procedure $\Sigma$ (generalized in
\cite{kerr2}) involves three successive transformations:

1) The electrostatic solution (real potentials $\E$, $\psi$, $\e^{2k}$) is
transformed to another electrostatic solution ($\hat{\E}$, $\hat{\psi}$, 
$\e^{2\hat{k}}$) by the SU(2,1) involution $\Pi$:
\be\lb{inv}
\hat\E = \frac{-1 + \E + 2 \psi}{1 - \E + 2 \psi}\,, \qquad 
\hat{\psi} = \frac{1 + \E}{1 - \E + 2 \psi}\,, \quad \e^{2\hat{k}} =
\e^{2k}\,. 
\ee
In the case of asymptotically flat fields with large distance monopole 
behavior, if the gauge is
chosen so that $f(\infty) = (1+\psi(\infty))^2$, then the
asymptotic behaviors of the resulting electric and gravitational
potentials are those of the Bertotti--Robinson
solution \cite{br}, $\hat{\psi} \propto r$, $\hat{f} \propto r^2$. 

2) The static solution ($\hat{\E}$, $\hat{\psi}$, $\e^{2\hat{k}}$) is
transformed to a uniformly rotating frame by the global coordinate 
transformation $d\varphi = d\varphi' + \Omega\,dt'\,, \,dt = dt'$,
leading to gauge--transformed complex fields $\hat{\E}'$, $\hat{\psi}'$,
$\e^{2\hat{k}'}$. 
While such a transformation on an asymptotically Minkowskian
metric leads to a non--asymptotically Minkowskian metric, it does not
modify the leading asymptotic behavior of the Bertotti--Robinson metric,
so that the transformed fields are again asymptotically Bertotti--Robinson.  

3) The solution ($\hat{\E}'$, $\hat{\psi}'$, $\e^{2\hat{k}'}$) is
transformed back by the involution $\Pi$ to a solution ($\E'$, $\psi'$,
$\e^{2k'}$) which, by construction, is asymptotically flat, but now has
asymptotically dipole magnetic and gravimagnetic fields. As shown in
\cite{kerr}, the combined transformation $\Sigma$ transforms the
Reissner--Nordstr\"{o}m family of solutions into the Kerr--Newman family.

The static ring wormhole solutions of the preceding section have two
distinct asymptotically flat regions $x \to \pm\infty$. To apply the
general spin--generating procedure $\Sigma$ to such wormhole spacetimes,
we must therefore select a particular region at infinity. e.\ g.\  $x \to
+\infty$, and gauge transform the static solution ($\E_0, \psi_0(x)$) to
\be\lb{gauge}
\E(x) = c^2\E_0 -2cd\psi_0(x) - d^2\,, \qquad \psi(x) = c\psi_0(x) + d
\ee
with the parameters $c^2 = 1/f_0({+\infty})$, $d = -c\psi_0({+\infty})$
such that $\psi(+\infty) = 0$, $f(+\infty) = 1$.  A
perturbative approach to the generation of slowly rotating ring wormhole
solutions \cite{ch} shows that the asymmetry between the two points at
infinity thus introduced is a necessary feature of spinning ring
wormholes.

To recover the spacetime metric from the spinning potentials ($\E'$, $\psi'$,
$\e^{2k'}$), we compute from (\ref{twist}) the partial derivative
$\partial_y\omega_{\varphi}'(x,y)$, which is a rational function of $y$, and
integrate it with the boundary condition $\omega_{\varphi}'(x, \pm 1) = 0$
(regularity on the axis $\rho = 0$). The resulting spacetime metric is
of the form (\ref{stat1}), (\ref{oblate2}) with the metric functions
\ba\lb{spin}
f' & = & |\Delta|^{-2}(1 - \Omega^2\rho^2/\hat{f}^2)\,f\,, 
\qquad f'^{-1}\e^{2k'} = |\Delta|^2 f^{-1}\e^{2k}\,, \nonumber \\
\omega_{\varphi}' & = & \Omega\nu^2(1-y^2) \left[ \frac{|\Delta_0|^2(1+x^2)}
{b^2\psi^2(\hat{f}^2-\Omega^2\rho^2)} -
\frac{\hat{f}^2\xi^2}{1+x^2} +\eta(\eta-\alpha\frac{b}{c}) \right] \,,
\ea
where $b = d+1$, $\hat{f}(x) = f(x)/b^2\psi^2(x)$, $\Delta_0(x) =
\Delta(x,y_0(x))$ with $1-y_0^2 = \hat{f}^2/\Omega^2\nu^2(1+x^2)$, and
\ba
& & \Delta(x,y) = 1 + \Omega^2\nu^2b\psi(\alpha^2b^2/2c^2 + \xi(1-y^2)) - 
i\Omega\nu b\eta\psi y\,,
\nonumber\\ 
& & \xi(x) = (1 + x^2)/2\hat{f} - \alpha^2b^2/2c^2\,, 
\quad \eta(x) = x + \alpha(2b-1)/c - \alpha/c\psi\,.
\ea

We can show that zeroes of $\Delta(x,y)$, corresponding to strong
Kerr--like ring singularities \cite{kerr2} of the stationary solution, are
absent if 
\be \lb{reg2}
\alpha bc >0\,.
\ee
This quasi--regularity condition which, in the cases $\E_0 = -1$ or 0, is
equivalent to assuming the static mass $M_+$ to be positive, can always be
satisfied by choosing the appropriate sign for the constant $c$ in
(\ref{gauge}). The metric (\ref{spin}) is of course still singular on the
rotating cosmic ring $x = y = 0$, with the same deficit angle
$\pi(\E_0\alpha^2 - 1)$ as in the static case.  The gravitational
potential $f'$ vanishes on the stationary limit surfaces $\hat{f}(x) =
\pm\Omega\rho$, where the full metric is regular. However the spinning
solution (\ref{spin}) is horizonless, just as the corresponding static
solution \cite{kerr2}. This spinning solution is by construction asymptotically
flat for $x \to +\infty$, but not for $x \to -\infty$, where the
asymptotic metric behaves as
\be
ds'^2 \simeq -l^{-2}\rho^{-2}(dt + (\Omega/4)(\rho^2+4z^2)\,d\varphi)^2 -
16\Omega^{-2}l^6\rho^4(d\rho^2 + dz^2) + l^4\rho^4d\varphi^2
\ee 
(with $l^2 = \Omega^2/4\hat{f}(-\infty)$). The study of geodesic motion in
the metric (\ref{spin}) shows that all test particles coming from $x \to
+\infty$ are reflected back by an infinite potential barrier at the
stationary limit, so that there is no loss of information  to $x \to
-\infty$.

From the asymptotic behaviours of the spinning metric and electromagnetic
field, we obtain the corresponding mass, angular momentum, charge, and magnetic
dipole moment,
\ba\lb{paras}
& M = (\nu\alpha/c)(b - 1 + \tau)\,, \quad  
& J = \nu\beta(M + \delta)\,,\nonumber \\
& Q = (\nu\alpha/c)(1 - \tau) \,, \quad 
& \mu = \nu\beta(Q - \delta)\,, 
\ea
with $\beta = \Omega\nu\alpha^2b^2/c^2$, $\tau = \beta^2c^2/2\alpha^2b$, 
$\delta = \nu c(1-\E_0\alpha^2)/3\alpha b$.
 
Can the values of these parameters correspond to
those of elementary particles? Combining the above values we obtain
\be
M^2 - Q^2 - a^2 =  \nu^2(\beta^2 - \E_0\alpha^2) - a^2\,.
\ee
The quasi--regularity condition
(\ref{reg2}) implies $\delta > 0$, so that $a^2 > \nu^2\beta^2$, hence the
inequality (\ref{part}) is satisfied for $\E_0 \geq 0$. The gyromagnetic
ratio
\be
g = 2 \frac{M(Q - \delta)}{Q(M + \delta)}
\ee
can be very close to 2 for very small values of $\delta$. One would then
expect that the values of the independent free parameters $\nu$, $\alpha$,
$\beta$, and $\tau$ can be adjusted so that the four physical
parameters (\ref{paras}) take their elementary particle values. However,
owing to the regularity constraint (\ref{reg1}) which restricts the range
of allowed parameter values, it turns out that if for instance the charge
and angular momentum are of order unity (in Planck units) and $g \simeq
2$, then the mass of these spinning ring ``particles'' should be at least of
the order of the Planck mass.

In the case of large quantum numbers ($|J| \gg m_P^2$), our classical
solutions (\ref{spin}) describe macroscopic closed cosmic strings with
negative deficit angle, but positive total mass. These cosmic strings
satisfy the elementary particle constraint (\ref{part}) if $\E_0 \ge 0$.
The exotic line source $x=y=0$ is spacelike only if it lies outside the
stationary limit surface, which further restricts the parameter values. A
specially interesting case is $\tau = 1$, corresponding to a neutral
spinning cosmic string ($Q = 0$). In this case we can show that the ring
source is spacelike for $|\beta| < 1/2$ (leading to $\delta/M > 4/3$); for
$\E_0 = 0$, the range of $|\beta |$ can be narrowed down to $0.29 <
|\beta| < 0.40$. These neutral strings have a proper perimeter of the
order of or smaller than their Schwarzschild radius $M = |\beta|\nu$, a
rotation velocity which is close to 1, and a magnetic moment $|\mu| \ge
\nu^2/3$, corresponding to a current intensity of the order of the Planck
intensity. The stability of these exact closed string solutions remains to
be investigated.

\end{document}